\documentclass{article}

\usepackage[english]{babel}
\usepackage[usenames]{color}

\begin{document}

\title{Introducing Sourcements%
\thanks{This work has been performed  in the context of the NWO project
Symbiosis which focuses on software asset sourcing.
The authors acknowledge  Karl de Leeuw, Sanne Nolst Trenit\'{e}, and
Arjen Sevenster (University of Amsterdam), for discussions concerning outsourcing. When referencing to this paper we will use the following author names with initials as follows: J.A.\ Bergstra, G.P.A.J.\ Delen and S.F.M.\ van Vlijmen.}
}

\author{%
  Jan Bergstra$^{1,3}$\\Guus Delen$^{2}$\\Bas van Vlijmen$^{1}$
\\[2ex]
{\small\begin{tabular}{l}
  ${}^1$ Section Theory of Computer Science,
  Informatics Institute, \\
  Faculty of Science,
  University of Amsterdam, The Netherlands.\\
  ${}^2$ Verdonck, Klooster Associates, Zoetermeer, The Netherlands.\\
   ${}^3$ Department of Computer Science,
 Swansea  University, UK.
   \end{tabular}
}
\date{}
}

\maketitle

\begin{abstract}
\noindent
Sourcing processes are discussed at a high abstraction level.
A  dedicated terminology is developed concerning general aspects of sourcing. The
term sourcement is coined to denote a building block for sourcing.  Notions of allocation, functional architecture and allocational architecture, equilibrium, and configuration are discussed. Limitations of
the concept of outsourcing are outlined.

This theoretical work is meant to serve as a point of departure for the subsequent
development of a detailed theory of sourcing and sourcing transformations, which can be a tool for
dealing with practical applications.
\end{abstract}

\section{Introduction}\label{sec:Intro}
The need for a theory of sourcing transformations from practice: given the abundance of sourcing transformation processes there is a need to
formulate professional standards for these.  Such standards can only be put in place when a theory about sourcing transformations is available. Clearly that theory must
be complemented with a vast amount of practical experience, for instance, consensus
about concepts must be obtained, and it must be rooted in practice. This paper is not reporting such a
consensus, but merely about the outline of a larger framework of thought in which such a theory of sourcing transformations might be embedded.

\subsection{Sourcing versus outsourcing}
Outsourcing is the most well-known sourcing transformation type and we will use ``theory of outsourcing''
as a shorthand for ``theory of sourcing transformations'' and in some cases ``outsourcing'' will be used as
a shorthand for ``sourcing transformation''. These meanings are secondary, however, with outsourcing denoting a specific (type of) sourcing transformation taken to be the primary meaning of the term.

In Delen \cite{Delen2005,Delen2007} a survey is given of definitions of outsourcing, insourcing, outtasking, intasking, follow-up sourcing, back-sourcing, greenfield outsourcing and greenfield insourcing.\footnote{As early references concerning sourcing Delen mentions: \cite{KernWillcocks1999}, \cite{Lacity1993}, \cite{LohVenkatraman1992}.} In addition to these notions multiple outsourcing has become common practice, a definition thereof can be found in \cite{Beulen2000} where multiple outsourcing is subsumed under outsourcing.%
\footnote{See \cite{WibbelsmanMaiero1994} for a relatively old source for this matter.}

In \cite{BV2010} it is argued that in the definitions surveyed by Delen and in many later
definitions ambiguities concerning the concepts of insourcing and outsourcing are built in.%
\footnote{%
In computer science, for some reason, many definitions are deficient. In \cite{Middelburg2010a} it is explained that the experimental process of software testing lacks a definition in the classical literature, and in \cite{Middelburg2010b} it is argued that even for the omnipresent concept of an operating system theoretical informatics provides no informative definition. %
}
\subsection{Outsourcing as a sourcing transformation}
An attempt has been made in \cite{BV2010} to disambiguate the concepts of insourcing and outsourcing. We will take that paper as the point of departure for further work in this paper. The central design decision of \cite{BV2010} is that outsourcing and insourcing  will refer to transformations of a state rather than to steady states existing before or arrived at after a state transformation. The relevant states are sometimes called sourcing states, or simply sourcings. Thus outsourcing and also insourcing will be understood as transformations of sourcings.%
\footnote{In \cite{ERPR2006} a survey of theories of outsourcing is given. Of the 16 theories surveyed 8
take outsourcing to be an act of change rather than a qualification for a state of affairs. In \cite{ERPR2006} an attempt is made to combine the disparate meanings that were found in these 16 theories. Their result is
the following:
\begin{quote}
``Outsourcing is a strategic decision that entails the external contracting of determined non-strategic activities or business processes necessary for the manufacture of goods or the provision of
services by means of agreements or contracts with higher capability firms to undertake those activities or business processes, with the aim of improving competitive advantage.''
\end{quote}
An obvious virtue of this definition is that it is not a simplistic one-liner. On the contrary, it tells a
specific story, and it is quite clear for that reason. That clarity also invites criticism. We will not follow this definition for several reasons:
(i) outsourcing is not a decision,
it may at best be the implementation of a decision,
(ii) the definition includes criteria for success (that the insourcer must be of a higher capability),
which must be left to external observers, at least in principle,
and (iii) the aim is too narrow because outsourcing can take place in the absence of competition.
}

The main argument for disambiguating the terminology that was put forward in \cite{BV2010} is that very clear meanings are needed if one intends to analyze, describe and define complex sourcing issues. Ambiguity of meaning by itself is unproblematic but the way for instance outsourcing is used in writings on
sourcing indicates the occurrence of frequent switches between different meanings within the same paper.  In \cite{BV2010} the phrase `sourcing equilibrium' has been used to denote a state comprising a number of units, each possibly decomposed into subunits, together with sources and belongs-to relations, as well as usage relationships between them.

The work in \cite{BV2010} suggests a need for more precision concerning the details of
particular sourcing transformations.  Providing a terminology that allows for a higher resolution is a
non-trivial task by itself, however, and we have concluded that it requires as a precondition
the availability of a more expressive terminology for sourcing phenomena in general.
The objective of this paper is to provide such a general context for a sourcing theory. This amounts to
 a fairly general story about the global concepts of sourcing and their interrelations. We will not attempt
to design more precise terminology concerning sourcing proper based on this general development
here, however.

Because detailed descriptions that may support a bottom-up development of sourcing theory are considered unavailable at this stage and because we favor a top-down development of the terminology, a naive understanding of the relevant notions, if necessary equipped with more detail from \cite{BV2010},
suffices as the basis of what follows below.

\section{Comparing sourcing to investing as terms}
Investing can refer to an individual purchase of specific assets by a specified agent at
a definite moment in time, based on given expectations, and involving known or at least
marginally analyzed risks. This description of investment focuses on how a
particular investment is made.

Investing may also refer to a process where a given amount of funds is spread over an
investment portfolio through a period of time. Investing a sum inherited
from a relative may take this form, and in this case the focus is on what is invested.

At a more distant level investing may indicate a method or strategy which is
applied by a unit, for instance a pension fund, if it receives a stream of funds.
The incoming stream of funds is grouped into blocks that are invested in the
previously specified sense of the term.

The process of investing, when successful, leads to so-called investments.
Those investments in turn can be assessed against the original purposes
of the investing units. Investing may be understood as funds allocation,
with an investment denoting a coherent part of (the result of) such an allocation.

The term investment has the following ambiguity: it may both refer to the (dynamic)
act of investing as well as to the (static) result of that act. If this ambiguity is potentially
problematic the static meaning is usually preferred, however,  together with a preference
for a dynamic meaning of investing, so that investing results in an investment. Many
different terms exist and (the act of) investing may also be referred to as a purchase, an acquisition, a transaction, or simply as an investment decision. Investing will be performed in order to profit
from the sale of the resulting investment in the future. Investing does not, in general, imply an
unbounded commitment to holding the resulting investment, by default it is rather the other way around.
Indeed, investing is done with future disinvesting in mind, though not necessarily with a planned
moment for doing so.

Sourcing as a term can be understood in a way comparable to investing.
Sourcing concerns the allocation of sources needed or used for producing
processes and services. Sourcing inherits the ambiguity of ``allocating'' which both denotes
a static ``state of having been allocated'' and the process that leads to that state. If
the process is meant, sourcing may also refer to a method for solving issues that reoccur with some frequency.

Sourcing may  involve just a specific source, or a package
of sources and it can also refer to a strategic approach to reoccurring sourcing issues.
These three meanings are similar to the three levels of meaning that were distinguished
for investing above.%
\footnote{An investment may be understood as a well-defined asset wrapped in a
well-understood context. That will hold for a source as well in general.}%

Sourcing, if understood as the process of sourcing, or rather of source allocation,
leads to a state. A concise
term for indicating that kind of state seems to be missing. We propose to
use the term {\em sourcement} for that purpose, in spite of the fact that this term
has different connotations close to the topic at hand.

A {\em basic sourcement} comprises only one source. Sources will not be
distinguished if they are always in the hand of the same owner and under
control of the same management.

\subsection{Sourcement: the need for an additional term}
When outsourcing a source is sometimes wrapped in a service. One may claim that each
source can in principle be wrapped in the service of providing that source. At first sight
the notion of a source seems to be the cornerstone of sourcing theory (sourcing science)
just as the notion of a service underlies service science. (We will discuss service science in more detail in Section \ref{service-science} below.) But this is wrong: sourcing is
intrinsically linked with units that serve as owners of sources. The simplest notion
of a (module of) sourcing is just that: a source assigned to an owner.
That is also the simplest
form of a sourcement. Sourcements, however are closed under composition, and for
that reason sourcements can have an architecture, can be engineered in a systematic fashion.  Importantly, sourcements must often be sufficiently comprehensive to enable an assessment of their business cases from different perspectives.

Having the term sourcement available, sourcing can be understood as the provision of adequate sourcements. Stated differently: the result of sourcing transformations, which may
result from the application of a sourcing strategy is a sourcement, or if one so prefers a sourcement portfolio. Of course adequacy of a sourcement portfolio is assessed in terms of
business objectives which at the same time are the drivers of achieving sourcing transformations and of developing sourcing strategies. Where service engineering and service provision lead to the being in place of a service, sourcement engineering and sourcement provision (that is sourcing) lead to the being in place of a sourcement.

Once equipped with the notion of sourcement, outsourcing can be understood as a sourcement transformation which is of a dominantly outsourcing character, that is:
\begin{enumerate}
\item\label{domoc1}
the outsourcing unit
discontinues ownership of some sources (disposes of some sourcements from its
sourcement portfolio) and,

\item\label{domoc2}
instead of the disposed sourcements new sourcements appear (in the
sourcement architecture after the transformation) that allocate equivalent or in some way
corresponding (though perhaps more, or in rare cases even less, effective) sources to other units.
\end{enumerate}

The conditions \ref{domoc1} and \ref{domoc2} do not define outsourcing because there is no guarantee that the outsourcing unit can keep its mission protected. We return to that matter later in more detail. Instead \ref{domoc1} and \ref{domoc2} merely serve the purpose of narrowing down the class of sourcement transformations that can be classified as outsourcings.

These considerations justify the use of sourcement as a basic term in the sourcing theory jargon, except for the fact that alternative terms or phrases might express the same intuition in a better way. For that reason we have listed some alternatives below, none of which, however, we consider superior to sourcement at least for the purposes just outlined.

\subsection{Alternative terms and phrases}
Several alternatives can be found for sourcement:
\begin{enumerate}
\item sourcing arrangement,
\item source allocation, (notice that source allocation may also denote the act of
allocating, while source de-allocation must denote a transition),
\item sourcing module (or: source allocation module),
\item source configuration (the process of
developing and maintaining source allocations may then be called
source configuration management),
\item sourcing configuration,
\item local sourcing equilibrium.
\end{enumerate}

We will assume that each of these phrases can be used as an alternative to sourcement.
Important concepts often have different wordings in order to allow for flexible language.

The proposal to equate these phrases with sourcement mainly implies that none of these
should be burdened with a clearly different meaning in the context of sourcing, at least not in
any further development of sourcing/outsourcing theory that takes this paper's  introduction of
the notion of a sourcement as a point of departure.

\subsection{Granularity of sourcement I}
The term investment exists besides the phrase investment portfolio. It is a matter of taste
to some extent from what point onwards an investment is preferably seen as an investment
portfolio. This is a matter of granularity. It would be quite uncommon to
consider each individual stock of a particular fund available on the stock market
as an individual investment and their
grouping together as an investment portfolio. Reasonable granularity
indicates that it is more plausible to consider the whole of stocks of the same fund or corporation
as a single investment. The phrase investment portfolio allows a modular approach
to investment analysis and management.

Similar consideration can be made about sourcements. Once sourcements have
been introduced as coherent source allocations ``sourcement portfolio'' emerges as a
plausible reference to  combinations of sourcements. Whether in a specific context there is a
need for that form of modularity, and where boundaries must be placed are matters
that must be dealt with by a fine-grained theory of sourcing.
At a coarse-grained level no more can be said than that sourcements
may be so extensive that it is more plausible to refer to them as sourcement portfolios.

\subsection{Granularity of sourcement II}
As it stands sourcing transformations involve human operation. That is each
successive step involves meticulous decision making. Computer support of
these decisions is plausible but full automation is not. If a unit makes use of
grid computing and the grid management system decides to run an executable
code on a remote machine instead of on its original host owned by the unit,
that is not an instance of outsourcing although it comes close. The difference is
that no human decision making process is taking place. Instead the entire
grid must be considered a single source and moving tasks around is a part
of its proper functionality.

Thus sourcing theory is about human centered, perhaps also computer supported,
decision making about changes that are not already covered by the functionality of an
existing source of a combination of sources, perhaps involving human employees who
perform fully specified tasks and take well-understood resource allocation decisions.
All these issues are relative to the
structure of a unit. Decisions about sourcing will by necessity involve the
top-management of that unit. If a unit's top-management is not involved in a source
allocation decision (e.g. hiring a van to transport some equipment for a unique occasion) then
seen from the level of the unit that
decision takes place as a consequence of normal operational procurement processes.

\section{Service Science: a competitor or an ally?}\label{service-science}
When contemplating the development of sourcing theory an obvious role model
is service science as it has been formulated by many authors since around 1995,
with IBM based researchers in a sustained leading role.

Not only is service science a role model for ``sourcing science'', it is in fact so
closely connected to the topic of sourcing that
one even needs to justify the need for a story about sourcing outside service science.
Indeed every source can be wrapped in the service of providing that source. Here we assume
that a service has an owner (producer, host). That fact is often left implicit but it
transpires clearly from the definition used in \cite{MaglioSpohrer2008} which clearly
speaks of a service
as a feature of a networked system where one or more systems are capable of
improving the state of another system. We assume that ``another'' refers to a real
form of identity which may and is even likely to involve differences of mission and objectives.
In addition we assume that complementary to another the self is not anonymous.%
\footnote{An anonymous self is usually assumed in the objects and threads created during
the run of a program written in an object oriented programming language.}
Now service science has a significant advantage over sourcing theory in terms of
its stage of development. Following the work of Vargo and Lusch starting with
\cite{VargoLusch2004} so-called Service-dominant logic (S-D logic) has been developed.
S-D logic clarifies how one might perceive service as dominant over goods. The dominance
of goods (G-D logic) is claimed to have been unchallenged from the early days of
economic science. But that is now changing.

Whoever adheres S-D logic is likely to consider sourcing a theme subsumed under
service science. In spite of the attraction of S-D logic we suggest that sourcements can
be considered dominant. This is simply the classical viewpoint that owners of the
means of production are the decisive factor. A source is a component of a
sourcement or of a service.
It serves as a means of production in processes with inputs and outputs
qualitatively differing from the source. S-D logic explains how services can be
dominant over goods, but the
role of goods is fully acknowledged. Indeed knowledge about services is an
addition to but not a replacement of knowledge about goods.

A sourcement-dominant logic may explain circumstances where sourcements are
dominant over services and still accommodate the crucial importance of a service
oriented perspective
when fruitful. We will not propose to replace S-D logic by Sourcement-dominant logic but it
might be an addition to it which leads to a more balanced view.%
\footnote{Sourcement-dominant logic places an emphasis on RBV, the resource-based view of the
firm as proposed by \cite{Wernerfelt1984}.}

\subsection{Sourcement-dominant logic}
Here are two examples where it is reasonable to attribute dominance of the sourcing perspective over the service (provision) perspective. With these examples in mind the subsequent listing of fundamental premises for an (as yet hypothetical) Sourcement-dominant logic can be appreciated.

\subsubsection{Highly skilled personnel}
In a professional sports team, or in an academic research institution, highly trained and
probably talented personnel may be viewed as sources. Together with a team affiliation
a member comprises a sourcement. Of course a skilled team player might be understood as someone offering a service to his/her colleague team mates. But that is a secondary perspective, the sourcement perspective, involving players who may have binding contracts
with their team managers,  perhaps with a duration of several years, is more convincing.

\subsubsection{Cloud computing}
Nowadays cloud computing is extremely fashionable. Obviously a user of a cloud uses a service and for that reason, seen from its users, cloud computing can be considered a
topic in service science.

In practice a cloud is implemented on a proprietary grid which contains one or more
highly efficient computing and data storage centers. Only parties who own massive low cost processing can offer the services expected from a cloud provider with competitive prices.
Thus, although the cloud is sold as a service or a package of services, it is based on the
availability of a very scarce source: the highly productive and efficient data processing and  storage center. In the context of cloud computing
the story of services simply hides (or even obscures on purpose) the
dominant economy of scale offered by centralized computing performed for a heterogeneous
group of clients in different parts of the world.

\subsubsection{Fundamental Premises of Sourcement-dominant logic}
Some fundamental premises of Sourcement-dominant logic can be listed:
\begin{enumerate}
\item
A source type is essential for a process or for a service if any architecture for that process or service features one or more sources of that type. A highly competitive unit that provides sources or processes needs to ensure that it makes use of most effective sources of necessary source types, even if that requires outsourcing.
\item
A sourcing strategy must allow a flexible method to adapt the sourcement portfolio when new instances of essential sources emerge inside or outside a unit.
\item
Sourcements containing highly efficient sources are in principle a candidate for being offered as a service to other units as well.
\item
Control over a highly efficient sourcement is always a candidate for being a significant
part of a unit's mission. (Even if the sourcement serves merely as a tool for its original mission.)
\item
Assuming that a unit is permanently optimizing (strengthening) the sourcement portfolio that it makes use of for the realization of its mission, then at any time its mission may switch to one in
which sourcement ownership and exploitation takes priority over previous missions.

If this instability is cast as  a step towards subscribing to a service dominant logic that
position does not affect the diagnosis that the quality of the sources has been driving the unit towards  that step.
\item
If a very effective sourcement is not used in a best possible way, in a competitive market the sourcement is likely not to be cost effective.
\end{enumerate}

\subsection{Sourcement usage versus service usage}
A unit may make use of a service offered and delivered by another unit. If this use is incidental
we do not speak of external sources when referring to the sources owned and made use of by the unit offering the service (for producing the service). The notion of a source involves an aspect of permanency
which is not required when use is made of a service.

Thus if mission critical sourcements are said to be outsourced%
\footnote{It is assumed that a sourcement is outsourced if all of its sources are outsourced.
}
in favor of the usage of a service hosted
by another unit, it may be the case that at a closer inspection these sourcements are simply discontinued
and a service contract is signed as a compensatory measure.

\section{Limits of (out)sourcing}
As a practice and field of expertise, sourcing has developed from the narrower practice
of outsourcing and its counterpart activity of insourcing. See for instance \cite{FreytagKirk2003} for an extensive systematic discussion on how outsourcing and insourcing fit in a company's strategy,
as well as the use of sourcing as a more general term. Sourcing lacks the directional asymmetry of
outsourcing and insourcing. And indeed, sourcing is nowadays often felt to be more general where as a field of expertise outsourcing (or even outsourcing plus insourcing) is considered to be too narrow for covering an experts professional identity.

However we will defend that outsourcing must not be subsumed under sourcing as a field of expertise, in spite of the fact that the activities subsumed under outsourcing, insourcing, backsourcing, and follow-up outsourcing may not constitute an autonomous field of expertise either.%
\footnote{If, however, one understands outsourcing as the state of having been outsourced,
or, even more  demanding as some
authors do, as (i) having been outsourced, while (ii) that goes against common practice, then outsourcing can be subsumed under sourcing. Our dynamic interpretation of outsourcing prevents this, however.%
}

Here are some essential aspects of outsourcing which are less prominent in sourcing:
\begin{description}
\item{\em Human activity based.} Outsourcing involves organizational change that requires specific
human supervision and management attention. It cannot be automated by definition.
\item{\em Mission sensitive.} Outsourcing is limited by considerations of unit mission, having to do with being mission (co)defining, a notion that will be explained below. Such considerations may imply that
transferring a sourcement to
another unit and turning its original use into the use of a service wrapped around the source
by its new owner may not qualify as an outsourcing.
\item{\em Relevance sensitive.} If the resulting service is made use of only incidentally but not as a normal part of the unit's operations it is more plausible to hold the view that the  sourcement's existence has been discontinued, perhaps in combination with it having been sold to another unit. Thus, outsourcing a sourcement makes sense only if it is used often, which is close to being mission critical. We use the
weaker term relevance and find that as a concept outsourcing is relevance sensitive.
\end{description}

Symmetrically insourcing features aspects outside sourcing such as: (i) what is the impact on the identity of the insourcing organization? (ii) must it sharpen its mission statement after insourcing? (iii) should it subsequently outsource what became side-lined as a consequence?, and (iv) must it be on the lookout for insourcing options in order to strengthen its core competence base?\\

The question why and when outsourcing is an option is best approached in terms of examples. Finding a general story that covers the examples is of secondary importance.

\subsection{(Non)-outsourceability by example}
For each organization or type of organization one may ask several questions about its potential for sourcing:
\begin{enumerate}
\item Which sources or activities can be outsourced? For instance:
\begin{itemize}
\item Can the quest for innovation of a core process be outsourced? (Yes, in an industry which is
dominated by capital investments, for instance the exploitation of a tunnel or of a railway; no, in
an industry where innovative capacity is itself is absolutely necessary for competitiveness, for instance
in mobile phone manufacturing.)
\item Can a student outsource reading the textbooks? (Yes, if the problem is to find some type of
quotes; no, if the objective is to prepare for passing an exam.)
\end{itemize}
\item Where are the limits of outsourcing? For instance:
\begin{itemize}
\item  A university may feel the need to perform research and teaching ``itself''.
\item Another university may be happy to outsource
all of its teaching activities but insist that examination is not outsourced.
\item Yet another institution of higher education may decide to outsource its research
activities and focus on teaching exclusively
\item  In particular if unit $U$ outsources its research activities then $U$ may still be active in fund-raising for research and in distributing the funds raised to researchers from other institutions.
\item A mountaineer can outsource track finding (by making use of the service of a guide), but (s)he
cannot outsource walking (by being carried around). It is generally considered
imperative that a mountaineer is
self-propagating once having arrived at one of the agreed  base points
of the mountain under attack.
\item Consider a household $H$ which operates quite independently from other households, though it needs to buy all its food. It is intuitively clear that the entire task (say $s$) of food preparation can be outsourced
from unit $H$ to some service provider $P_s$, say another household. Whether or not that is economically feasible is another matter, but it can be done. On the other hand consuming the food cannot possibly
be outsourced. However obvious this may be, it indicates that some processes and activities
can be outsourced in principle whereas other are not capable of being outsourced.
For this reason each sourcing strategy needs to depart from a clear mission statement for a unit.%
\footnote{%
In \cite{FreytagKirk2003} a diagnostic tool is presented which centers around determining a company's mission. This method supports so-called continuous strategic sourcing.} It must be known which activities or sources must
a unit preserve because outsourcing those leads to a breach of the mission statement. Such
sources by definition cannot be outsourced.
\item
In \cite{King2007} it is suggested that a company may have informational core competences
which may reside in the development process of mission-critical systems.  This ability will probably not be outsourced according to \cite{King2007}.%
\footnote{
In \cite{HHB2010} another conclusion is drawn, however. Mission critical software for a
specific unit is likely to be less amenable for commodity outsourcing and specific methods are needed to achieve best of breed multi-vendor outsourcing arrangements. The paper provides a graphic method for visualizing the temporal evolution of such arrangements.}
\end{itemize}
\end{enumerate}

\subsection{Mission co-defining sourcements can't be outsourced}
Summarizing what was presented above the following is obtained: a sourcement is mission defining for a unit if outsourcing it cannot be done without adapting the unit's mission, and if in addition it is the only source owned by the unit with that property. A sourcement is mission co-defining if outsourcing it cannot be done without adapting the unit's mission and if it shares that privilege  with one or more other sourcements.

Mission (co-)defining sourcements are likely to be mission critical as well,
but not conversely. Needless to say that for an application of the criterion of being non mission (co-)defining, for not outsourcing a source, a unit must know its own mission,
and only then it can be determined which
of its sourcements are mission (co-)defining as well as mission critical.

Whether the design and engineering of sourcements which are mission co-defining and mission critical, involves any informational core competences is yet another matter. For the questions
we have just posed what matters is more focused, however: is the ability to develop a certain
mission-critical system itself part of the mission to the extent that it is mission co-defining,
only then outsourcing that competence is to be
rejected as a matter of principle because of the threat it poses to the unit's identity.

Here we find a conceptual difficulty: some sourcing transformations may fail to be labeled as outsourcing merely because the sources involved are mission (co-)defining. But that is a subjective criterion. It becomes possible that the concept of outsourcing is variable due to changes in a unit's vision concerning its mission. We have no clear response to this matter otherwise than to insist that the very concept of outsourcing is highly dependent on unit identities, and that an identity is more than a mere code or name. It involves a vision and a mission statement as well. For this reason the notion of outsourcing is
conceptually more demanding than the mission neutral notion of sourcing.

\section{Architecture}
Computer science shows two major uses of the term architecture: computer architecture,
which is about how the physical components of a computer are put together and software architecture which concerns the composition of software components. When both aspects are intertwined one often
speaks of systems architecture. The basic building block of architectural terminology in computing is usually a component.

\subsection{Systems architecture for sourcing}
Systems architecture which plays a role in sourcing combines aspects of both lines of
thought.  One may imagine an architecture of a unit or of group of units and sourcements therein
as:
\begin{itemize}
\item
A description of a number of sources giving information about their units and about
communication channels and use relations between them.
\item
Such a description is normative in the sense that it may be understood as a specification of
what the designer intended to build, rather than as a mere anatomy based snapshot of observed structure.
\item
Such a description usually predates the artifacts it describes (though that need not always be the case: reverse architecting produces a hypothesis concerning what was intended when the artifact was produced).
\end{itemize}
What remains of such an architecture if the information about units is forgotten? The most plausible answer is that a basic source after forgetting all units becomes a component of the more abstract architecture.%
\footnote{When in a basic sourcement the unit is forgotten (removed, abstracted from) simply a source
results. That source, however, plays the role of a component in the corresponding  unit-less architecture.}
Stated differently the link between conventional stories about architecture and architectural stories
concerning sourcing is that a sourcement is a component equipped with a unit serving as its owner and having control over its actions (at least at some level of abstraction).

\subsection{Architecture based sourcing}
A number of questions can be posed concerning the role that architecture may play for the solution of sourcing problems.
\begin{enumerate}
\item
Assuming that some specific theory of systems architecture is especially effective for
sourcing, one may ask what theory results after forgetting the difference of units. Given the vast
literature on architecture in computing, it is hard
to believe that a novel theory of architecture is found that way, and for that reason one may
turn the question around: which theory of systems architecture can be best expanded with the phenomenon of units as to obtain an architectural method suitable for dealing with sourcing problems?
\item
Another question is: given an architectural design of a system, how and why can it be
the case that its realization may be best distributed over a number of different units?
\item
To what extent is a sourcing problem (that is the problem to find an adequate sourcing plan for some unit or for a part of a unit) for some specific period given the unit's mission and targets primarily an architectural problem, with the distribution of sources over various
units being only of secondary importance?
\end{enumerate}

In any case, if a sourcing transformation is designed, some architectural task needs to be performed
assuming that the transition is to be realized in a planned way and preferably also in an evidence based fashion. What seems to be rather open is to what extent this task needs to
be based on modern methods of hardware and software architecture. In small and
medium size enterprises
a formalized and strictly methodological approach to such architectural tasks seems not
to prevail at this moment. In very large organizations that may be different.

\subsection{Architecture won't solve all problems}
A difficulty with architectural techniques taken from software engineering or computer
architecture is that one cannot expect these techniques to produce predictions on how
a business will proceed after an intended sourcing transformation has been brought
about by means of a planned transition. While architectural concepts are definitely useful for
specifying where to go, these may be less helpful when it must be judged why a certain transition is profitable, or even if its feasibility must be assessed. For this reason it cannot be claimed that
architecture is the intellectual core of sourcing to the same extent as that might hold for hardware or software design.

\subsection{Sourcing patterns}
In software engineering software patterns have become very popular. Making an inventory of
sourcing patterns might be as useful as an inventory of software patterns has proved to be.
Only serving as indicative  examples some sourcing patterns are mentioned below.

\begin{enumerate}
\item
The most prominent sourcing pattern is that of a shared service center within a unit.
\item
A computing grid based cloud.
\item
A third pattern is a competence pool. In a competence pool personnel with a specific range of competences is wrapped in a single service offered by an external unit.
\end{enumerate}

\section{Transformation and transition}
A sourcing transformation (or sourcement transformation which is the same) may involve a source removal, a source introduction, a source allocation (to some unit), a source de-allocation (from some unit), and more generally steps that result in source re-allocation.

The process that achieves a transformation is called a transition. A transition follows a plan which is like an algorithm. Its design as well as its execution is performed by human
operators, and these operators may be found in yet another unit which disappears from the scene once the transition is done. Planning and performing the transition, given the specification of a sourcement transformation, is itself a task which seems to induce a sourcing issue. That is misguided, however, because sourcing is about enabling methods for
primary processes of units and enacting its own transition is not one of any units primary processes.

Enacting the transition of other units can very well be among a unit's primary processes and
in that case parts of that work may be outsourced. When for instance, another unit is asked to provide
part of the work on a sustained basis, that may be a proper case of outsourcing.

Anyhow, if consultants or legal experts are hired to plan and perform a transition in order to achieve an intended sourcement transformation, such forces are most plausibly not seen as sources, because of their incidental involvement.

Because sourcing transitions are effected in a planned fashion by the action of a human organization, it is plausible that a transition itself moves through a formalized  life-cycle. In fact each unit source relation percolates through some life-cycle, which may of may not have been made explicit. For an informal discussion of life-cycles we refer to \cite{BergstraVanVlijmen1998} and for an instantiation of a life-cycle model for sourcing we mention the PON sourcing-cycle of \cite{PON2006}. The description of a sourcing transformation may be considered a control code in the sense of \cite{BergstraMiddelburg2009} and the organization that carries out its realization serves
as, or may be compared to, its executing machinery.  If different sourcing transitions are active in parallel the risk of unexpected and unanalyzed feature interaction (see for instance \cite{KimblerBouma1995}) becomes high.

Carrying out transitions correctly is often so complex that it obscures the (architectural)
design problem that needs to be solved when reallocating sources. In particular
governmental units required to apply the European Tender Process may feel themselves
compelled to pay more attention to the correct executability of a highly formalized tendering process than to finding the best possible solution to their sourcing problems.

\section{Equilibrium}
When considering a community of hundreds or even thousands of units it is very likely that sourcement transitions take place all the time. Thus an equilibrium, understood in terms of sourcements being fixed and stable, is unlikely to exist beyond some critical size (or below some critical size depending on the kind of equilibrium one is looking for). For a smaller unit, or for a smaller subunit of a large unit, it is likely, however, that its sourcement or sourcement portfolio remains stable (that is unchanged) for extensive periods of time. That situation will be referred to as a local sourcing (or sourcement) equilibrium.

A global sourcing equilibrium is likely to exist only if some abstraction is made. In a Walrasian equilibrium for micro-economics it is prices and trade volumes rather than ownership or distribution of assets which is stable during some period. For a global sourcing equilibrium to make sense some abstraction must be found, for instance the fraction of sources of a specific type which are positioned outside the unit whose primary process depends on these sources, or the fraction of sourcing transformations which are outsourcing transformations. Of course as
outsourcing can be seen as an economic transaction average prices of the resulting service deliveries, or of the average cost of making a transition, may also be used to obtain a level of abstraction which admits for a valid definition of equilibrium states.

At the micro scale, where individual units or subunits are considered we have microscopic sourcing theory. At the micro level one imagines that the sourcement portfolio of a (sub)unit
moves from local equilibrium to local equilibrium while transition phases are sequentialized and last short in comparison to the equilibrium phases. But even at the level of a unit for which further decomposition into subunits is unreasonable because of its coherent management structure it is quite conceivable that several sourcing transitions are carried out concurrently.
Thread algebra as formulated in \cite{BergstraMiddelburg2007} may be used as a conceptual model for the form of multi-threaded concurrency which takes place when the same unit is
executing several transitions in parallel.

Working towards an equilibrium can be taken as the conceptual point of departure for system administration (see \cite{Burgess2007}). System administration is related to IT sourcing, though it need not exclusively consist of man powered actions.

\section{Political economy}
Sourcing as a social phenomenon deserves its own historiography. We will not try to
write it here, but we put forward that a trend towards outsourcing marks the beginning of a
period in which many organizations need to have explicit and competitive sourcing policies.

In Delen \cite{Delen2005} one finds that an outsourcing decision requires that both
outsourcers and insourcers have an economic interest in the deal, and even stronger
they need to have a business case rationalizing that interest and they need to know and to
understand each other's business case.%
\footnote{This requirement is quite strong and it must perhaps be relaxed to requiring that outsourcer and insourcer both understand why the deal complies with the each other's missions.}
As a social phenomenon this requirement may be understood as the hypothesis that the outsourcing trend finds its origin in the development of
novel ways for doing business in a mutually profitable way. We will put forward a rather different motive with roots in the analysis of political economy due to Marx.

Rather than viewing Marx as the instigator of failed models such as the communism of the previous century, he may be considered a sharp observer of the emerging capitalism. In that quality
he has been and still is very visible in economics.

Marx assumed that an original split (separation) between labor and capital
(\emph{Urspr\"{u}ngiche Trennung} see \cite{Bonefeld2010}) prepares a setting where
workers produce added value
(\emph{Mehrwert}), which in fact can only be measured by means of money, and where the
capitalists, that is the owners of the means of production, accumulate the capital gains that accrue from taking the added value created by the workers' efforts. Marx thought that not only workers should be entitled to a significant share of the added value, but that in addition they ought to be freed from (i) the dependence resulting from not being owners of the means of production needed for their work, and from (ii) the degradation of their working conditions including a loss of perspective on autonomous and creative self-realization which he expected to result from the mentioned dependence.
\footnote{This is a perplexing view by any means because he also considered the original split part and its dismantling of a necessary and inevitable development, to the extent that he would applaud actions taken towards its demise taking place. He thought that only after the split the worker would develop the efficient methods of working needed to effect the progress from which from which more mature stages of development might then be obtained.%
}

We will not focus attention on these pessimistic expectations concerning working class living conditions. Remarkably capitalism has shown a formidable ability to avoid the Marxian
predictions to come true.\footnote{Many works have been devoted to this matter in a great variety of forms. We mention \cite{Klein2007} as an example.}
In fact and equally remarkably in the century after Marx's death so-called socialism and its close
relative communism came closer to the realization of the predicted ``Verelendung''.%
\footnote{A rather sceptical assessment of the intellectual legacy of Marx can be found in \cite{RagerRampeltshammer2005}. A suggestion that Marx may still constitute an incentive for current research can be found in \cite{Wolf2000}. These papers indicate that all official Marxisms are now gone and a  part of history. At the same time it is clarified that Marxism cannot exist anymore as a coherent philosophy because Marx' writings show too many inconsistencies. A most important inconsistency concerns his view that  farmers must become free in two simultaneous senses: free as citizens, and free from the burden of ownership of the means of production (that is the land). This dual freedom is necessary for the members of the working class in order to focus entirely on the improved organization of productive forces which in turn will be a precondition for overcoming capitalism in favor of socialism. These views have been misused in the period between the first and second world war in Russia. However, as \cite{Henniger2010} points out, in his later years Marx understood very well that the organization of agricultural sector in Russia's rural areas was worth a much more positive appraisal than he was aware of when writing his major works. Moreover, he explicitly warned against a simplistic application of his earlier views in that setting. Unfortunately this latter view had much less impact than his earlier position.}\newcounter{Marxfootnote}\setcounter{Marxfootnote}{\value{footnote}}

The critical point for our topic is the observation that systematically factorizing out similar functionalities from a range of different but comparable economic processes can form the basis of the construction of capital. So if some kind of task or activity can be extracted from a large number of units and can be combined together into a new unit, that unit can, after some further innovative development, represent new value and it can become a locus of capital accumulation.%
\footnote{Providing systems software for a PC has been an example of the remarkable capital accumulation that may accrue from the isolation of a seemingly marginal task. This observation does not presuppose that the factorization of working processes must lead to the
formation of a novel class (of so-called capitalists) who are exploiting some new form of proletariat.}

\subsection{Outsourcing as a driving force of innovation}
The crucial point is that seemingly simple or modest tasks when factored out from a sufficiently large number of units can be grouped together into new units which can grow into spectacular accumulations of sophistication, effectiveness and ultimately of capital.

If one assumes that the potential for accumulation of capital can itself be a driving force it becomes clear that insource business cases may be quite weak from outsourcer's perspectives, thus potentially compromising one of the rules laid down in \cite{Delen2005} (rules that are called here: decision and control factors). This may sound obscure as if the accumulation of capital is a primary objective which governs the development of business architecture. That may be disputed at first sight, but it cannot be denied that by building a  high profile and capital intensive unit around a collection of modest  but highly comparable processes that have been insourced from a wide range of outsourcing units these activities are much more likely to undergo innovation and technological evolution. Thus if for instance
university time-tabling is felt as a weakness, one can imagine that some time after its has been outsourced to a shared service and competence centre by and for a number of institutions it
can become so sophisticated that it turns into a recognized strength for the participating (outsourcing) units (institutions for higher education)
rather than remaining to be considered the problem it constituted before. Once that is true the value of the shared service centre explodes and it can be brought to the market in such a way that original owners make unexpected profits.%
\footnote{Thus what Marx understood as a mechanism that gives rise to class formation, can be understood as a necessary prerequisite for innovation. Marx understood the transition to capitalism through the division of work and ownership as a necessary transformation, a precondition for better times to come.}

\subsection{`Freedom in two senses' revisited$^{\arabic{Marxfootnote}}$}
The process of removing ownership of means of production from owners thereby obtaining
a pure work force as a result, may involve actions that are less than friendly towards the
original workforce. Anyhow, the idea is that freedom in two senses is achieved: freedom
from the ownership of the means of production (thus enabling a worker to focus on primary objectives and to participate in the strengthening of the workforce, nowadays called innovation), and freedom as a citizen. The second form of freedom has been achieved in most liberal democratic societies, and it ceases to play a role in describing transformations of labor division. Another way to understand this transformation is that the farmer is transformed from someone who sells food to someone who sells his or her own time and energy, and perhaps creativity.

We return to the time-tabling example from the previous section. After outsourcing, the workforce of time tabling staff is free from the obligations imposed by the overall tasks,
responsibilities, and objectives of an academic institution.
Instead their focus has become much more limited. At
the same time they are now in a much more competitive field an they need to work towards a sustained improvement of working procedures and cost effectiveness. After some time a sense of pride that was originally coupled with being a staff member of a reputed academic institution, and got lost during the outsourcing transition, may be regained, or rather replaced
by an awareness of being very competent and effective in a task which is mission critical for a
number of different academic institutions at the same time.

\subsection{Outsourcing as a unique driver of innovation}
A most amazing aspect of outsourcing is that activities which are considered of lesser importance than primary processes of a unit are among its most obvious candidates for outsourcing, and for that reason (after successful outsourcing) exactly those processes may become the locus of concentration and of subsequent innovation after which they may
emerge as drivers of further development in unexpected ways. So one concludes that not only
an expected cost advantage may trigger outsourcing of some type of activity but also a common awareness of a number of units of the circumstance that within their organization this particular type of activity fails to achieve the innovative power which it might have achieved in a different organizational structure.

\section{Concluding remark}
We have discussed sourcing in a general context, taking notice of its possible role in social structures at large. A proposal has been made concerning a modular terminology
for sourcing states, in particular the term sourcement has been coined for that purpose.
The role of architecture for sourcing issues has been discussed and several questions
about it have been raised though left unanswered. We consider this work to constitute a necessary element for the development of a theory of outsourcing. Such a development may, however, proceed in various directions and preservation of degrees of freedom for that development has been an
important objective for us as well.

\end{document}